\documentclass[aps,pra,twocolumn,preprintnumbers,nopacs,superscriptaddress,amsmath,amssymb]{revtex4}
\usepackage{dcolumn}
\usepackage{bm}
\usepackage{amssymb}
\usepackage[german, english]{babel}
\usepackage{graphicx}
\usepackage{color}
\usepackage{textcomp}
\usepackage[letterpaper,total={7in,9.5in},top=0.75in,left=0.75in]{geometry}
\usepackage{array}

\def\EE{\mathcal{E}}

\begin{document}
\title{Re-evaluation of the Beck \textit{et~al.} data to constrain the energy of the $^{229}$Th isomer}

\author{Georgy A.~Kazakov}
\affiliation{Vienna Center for Quantum Science and Technology (VCQ) and Atominstitut, TU Wien, 1020 Vienna, Austria}
\author{Thorsten Schumm}
\affiliation{Vienna Center for Quantum Science and Technology (VCQ) and Atominstitut, TU Wien, 1020 Vienna, Austria}
\author{Simon Stellmer}
\affiliation{Vienna Center for Quantum Science and Technology (VCQ) and Atominstitut, TU Wien, 1020 Vienna, Austria}

\date{\today}

\begin{abstract}
The presently accepted value of the energy splitting of the $^{229}$Th ground-state doublet has been obtained on the basis of undirect gamma spectroscopy measurements by Beck \textit{et al.}, Phys.~Rev.~Lett.~{\bf 98}, 142501 (2007). Since then, a number of experiments set out to measure the isomer energy directly, however none of them resulted in an observation of the transition. Here we perform an analysis to identify the parameter space of isomer energy and branching ratio that is consistent with the Beck \textit{et al.} experiment.
\end{abstract}

\maketitle

\section{Introduction}

The isotope $^{229}$Th possesses a nuclear isomer with an extremely low energy of only a few eV. A number of experiments have found evidence of the existence of this state, culminating in a direct detection experiment performed by the LMU group \cite{Wense2016ddo}. While indirect gamma spectroscopy measurements of its energy $E_{\rm is}$ have improved over the past 40 years (see Ref.~\cite{Peik2015ncb} for a recent review), direct measurements are not yet available. The commonly accepted value of the isomer energy is 7.8(5)\,eV \cite{Beck2007eso,Beck2009ivf}. This value has been verified by the LMU experiment, which contrained $E_{\rm is}$ to the interval between 6.3\,eV (first ionization threshold) and 18.3\,eV (third ionization threshold).

A number of recent experiments set out to measure the isomer energy by means of optical spectroscopy. These measurements include optical excitation of surface-adsorbed nuclei \cite{Yamaguchi2015esf} and of nuclei doped into bulk crystal material \cite{Jeet2015roa}, and detection of the isomer gamma emission following surface implantation of nuclei \cite{Zhao2012oot}. The failure to observe an optical signal can be explained in three ways:~(i) rapid quenching of the isomer through internal conversion processes, (ii) the isomer energy is outside the search range of the specific experiment, or (iii) the isomer lifetime is orders of magnitude shorter or longer than the expected value of about 1000\,s; see also \cite{Tkalya2015rla} for a recent treatment.

The failure of the UCLA (search range 7.3 -- 8.8\,eV) \cite{Jeet2015roa} and PTB (search range 3.9 -- 9.5\,eV) \cite{Yamaguchi2015esf} experiments to observe the isomer transition within the expected uncertainty range (7.3 -- 8.3\,eV) led us to revisit the original Beck \textit{et~al.} data to construct a confidence region for the isomer energy \footnote{The preparation of derivative works based upon original work that is protected by a copyright clearly is a copyright infringement. Prior to our work, S.~St. has obtained a license of APS (No. 3847551196238) to re-use Fig.~2 of Ref.~\cite{Beck2007eso}, including publication on non-profit websites.}.

\begin{figure}[b]
\includegraphics[width=\columnwidth]{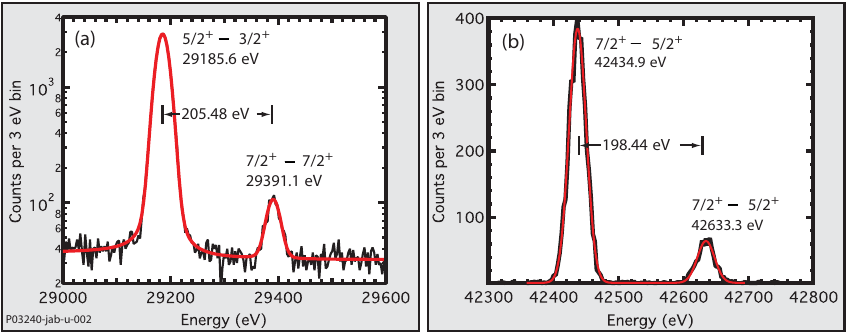}
\caption{The original data, taken from Ref.~\cite{Beck2007eso}.}
\label{fig:fig1}
\end{figure}

A related analysis had been performed by S.~L.~Sakharov in Ref.~\cite{Sakharov2010ote}. The author concentrates mainly on earlier indirect measurements, and shows that the obtained results strongly depend on the model of the decay pattern used to interpret the data. The study allows us to exclude all results of indirect measurements of $E_{\rm is}$ obtained before the Beck \textit{et~al.}~measurement from our considerations. Sakharov's criticism of the value $E_{\rm is}={\rm 7.6 \pm 0.5~eV}$ is based on two foundations:~underestimation of the error connected with the measurement of the position of the weak 29.39-keV line (Sakharov estimates it as 1.3 to 1.5\,eV), and possible corrections to $E_{\rm is}$ due to a different value of the 29.19-keV branching ratio $b$. However, these re-estimations have been done using some general considerations about properties of Gaussian fits, without the investigation of relevant experimental spectra. Here we perform a refit of the actual data.

The original publication \cite{Beck2007eso} makes very clear that their measurement is not capable of measuring the energy $E_{\rm is}$ of the isomer directly, but only the values of $\Delta E_{29}$ and $\Delta E_{42}$. Deriving the isomer's energy requires knowledge of the branching ratio $b$ of the 29-keV state to decay into the ground state, $E_{\rm is} = (\Delta E_{29} - \Delta E_{42})/(1-b)$. Obviously, the $E_{\rm is}$ heavily depends on the value of $b$, as emphazised in Ref.~\cite{Tkalya2015rla}. To illustrate the impact of $b$, we give a few examples:~$b=1/13$, as assumed in Ref.~\cite{Beck2007eso}, leads to $E_{\rm is}=7.6$\,eV, whereas $b=0.25$ gives $E_{\rm is}=9.3$\,eV, and $b=0.51$ gives $E_{\rm is}=14.0$\,eV, as calculated in Ref.~\cite{Sakharov2010ote}. In the present work, we show that this simple scaling is not compatible with the Beck \textit{et al.} experiment.

\section{Data extraction}

We use the data contained in Fig.~2 of Ref.~\cite{Beck2007eso}; reprinted in Fig.~\ref{fig:fig1} here. The figure comes as an uncorrupted vector graphics file, which allows for the extraction of the coordinates of all data points with nearly arbitrary resolution. We extracted the two coordinates of each data point with a precision of 8 digits, which is far better than required.

The coordinates are scaled by calibration with the respective $x$- and $y$-axes. For the $y$-axis (``Counts per 3 eV bin''), we benefit from the fact that the values are integer numbers:~the extracted value is rounded to the nearest integer, where the difference to the nearest integer is at most 0.02 counts. The 14 data points above 500 counts in the 29.18-keV peak (Fig.~\ref{fig:fig1} (a)) pose a bit of a problem, as they cannot be assigned unambiguously to an integer. We speculate that this specific sub-set of the data was processed in a way that resulted in non-integer values only in the peak of this feature. Alternatively, the logarithmic plot was generated with insufficient resolution. Whatever the cause, our values for these 14 specific data points are off by at most 2 counts per 3 eV bin.

As descibed in Ref.~\cite{Beck2007eso}, 10 calibration lines are used to ``stretch'' the energy axis near the 29-keV lines. The correction factor, as given in the text, is 0.999\,527(54), yielding a corrected bin width of 3.001\,42\,eV. From our analysis of Fig.~\ref{fig:fig1} (a), we extract a value of 3.001\,48\,eV, which is in good agreement. For the 42-keV data set, the calibration is poorly described, and the correction factor is not given. From the data set, we extract a value of 3.000\,14. We note that, without apparent reason, the correction term is exactly a factor of 10 smaller compared to the 29-keV data set. The applied ``stetching'' of the energy axis changes the derived value of $E_{\rm is}$ by about 0.1\,eV. Note that for the 42-keV doublet (Fig.~\ref{fig:fig1} (b)), data of only one of 25 high-resolution pixels is available.

\subsection*{The 29.374-keV line in $^{237}$Np}
\begin{figure}[b]
\includegraphics[width=\columnwidth]{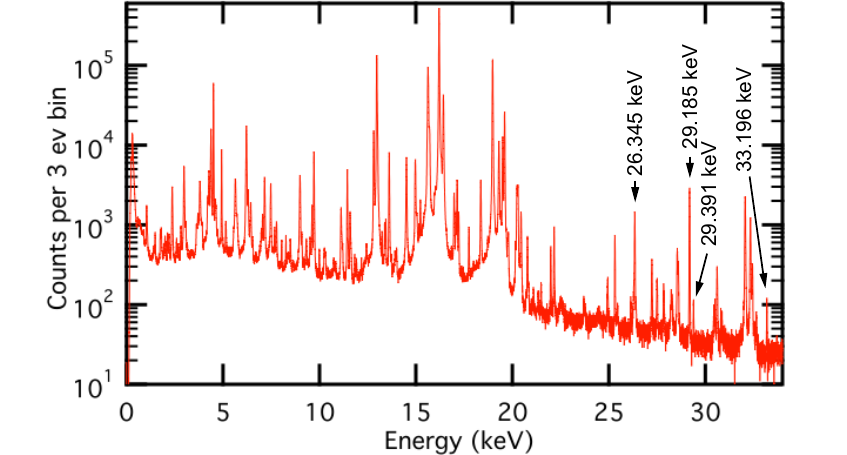}
\caption{The spectrum used for calibration; image taken from Ref.~\cite{Chodash2012talk}.}
\label{fig:fig2}
\end{figure}

The experiment uses an $^{241}$Am source for calibration, where the source is applied for about a quarter of the measurement time; see Fig.~\ref{fig:fig2}. The $^{241}$Am decays into $^{237}$Np ($t_{1/2} = 2.14 \times 10^6$\,a), which has a gamma emission line at 29.374\,keV. If strong enough, this line could perturb the 29.391-keV line of interest significantly. Such a contribution would render the observed isomer energy smaller than it is.

From the spectrum shown in Fig.~\ref{fig:fig2}, we extract the the position and amplitude of various lines. The energy uncertainty is less than 10\,eV for all lines, allowing for an unambiguous identification. The uncertainty in the amplitude is less than 10\%. We use two $^{241}$Am lines at 26.345\,keV and 33.196\,keV to ``sandwich'' the hypothetical $^{237}$Np line. After adjusting for the probability of these lines ($^{241}$Am(26.3\,keV):~2.27\%, $^{241}$Am(33.2\,keV):~0.13\%, $^{237}$Np(29.4\,keV):~14.1\%), and assuming the age of the $^{241}$Am source to be 50 years, we calculate that a peak caused by $^{237}$Np would amount to an amplitude of 0.15 counts per bin. This is much smaller than the actual height of the observed peak (about 100 counts). 

In addition, none of the other lines of $^{237}$Np (at 8.22\,keV (9.0\%) and 13.3\,keV (49.3\%)) and its daughter $^{233}$Pa (at 13.6\,keV (43\%)) could be observed. A disturbing effect of $^{237}$Np contributions can therefore be excluded.

\section{Confidence region based on the $\Delta E_{29}-\Delta E_{42}$ approach}

\subsection*{The LLNL experiment}
The experiment of Beck \textit{et al.}~measured the splitting between two pairs of energetically close $\gamma$-transitions, namely the (29.18, 29.39)~keV pair and the (42.43, 42.63)~keV pair, following the $\alpha$-decay of ${\rm ^{233}U}$. It employed a NASA X-ray microcalorimeter spectrometer \cite{Porter2004tae,Stahle2004tng} with an instrumental resolution of about 26\,eV. In this case (neglecting out-of-band branching ratios), $E_{\rm is}\approx (E_{29.39}-E_{29.18})-(E_{42.63}-E_{42.43})=\Delta E_{29}-\Delta E_{42}$. 

The value of $E_{\rm is}=7.6 \pm 0.5$~eV stated in Ref.~\cite{Beck2007eso} has been obtained by fitting of all four peaks by four lines, where $\Delta E_{\rm 29, meas}=205.48 \pm 0.50$~eV and $\Delta E_{\rm 42, meas}=198.44 \pm 0.22$~eV were obtained, yielding $\Delta E_{\rm 29, meas}-\Delta E_{\rm 42, meas}=7.0 \pm 0.5$~eV. This value has been corrected to $7.6\pm 0.5$~eV by taking into account the interband branching ratio $b(29.19 \rightarrow {\rm g})=1/13$, and to $7.8\pm 0.5$~eV by taking into account another branching ratio $b(42.43 \rightarrow {\rm is})$, where ``g'' and ``is'' denote the ground and isomer states of the ${\rm ^{229}Th}$ nucleus, and $29.19$ and $42.43$ denote the energy levels of this nucleus with energies given in keV. 

We note that the data extracted from Ref.~\cite{Beck2007eso} does not allow us to make a well-grounded conclusion about the correctness of the energy calibration procedure, the same is true for the splitting of the 42-keV doublet. Instead, we perform the fit of the 29-keV doublet using only the results provided in Refs.~\cite{Beck2007eso,Beck2009ivf} concerning the measured separation of the 42-keV doublet ($\Delta E_{\rm 42,meas}=198.44 \pm 0.22$~eV) and the branching ratio $b_{42}=0.02$ of 42.43 keV state into the isomer state. This branching results in a correction
\begin{equation}
\Delta E_{42}=\Delta E_{\rm 42,meas}-E_{\rm is} b_{42}.
\label{eq:2}
\end{equation}

The data for 29-keV doublet is presented as a set of $N=201$ pairs $(\EE_i,n_i)$, where $\EE_i$ and $n_i$ are the mean energy and the total number of counts per $i$th bin, respectively. The number $n_i$ of counts per $i$th bin is a Poissonian random number with an (unknown) mean $\lambda_i$. We can parametrize all these means $\{\lambda_1,...,\lambda_N\}$ by a model profile depending on the set of fit parameters $\theta=\{\theta_1,...,\theta_p \}$: $\lambda_i=\lambda_i(\mathbf{ \theta})$. The values of $\theta$ can be estimated using the maximal likelihood method. This method builds on the maximization of the so-called {\em likelihood function} $L(\mathbf{n}|\mathbf{ \theta})$ defined as a probability for realizing the experimentally observed set $\mathbf{n}=\{n_1,...,n_N\}$ at given values of the data. It is also convenient to introduce the {\em logarithmic likelihood function}
\begin{equation}
\ell(\mathbf{n|\theta})=\ln L(\mathbf{n|\theta})= \sum_{i=1}^N \left[n_i \ln \lambda_i(\theta)-\lambda_i(\theta)-\ln(n_i!) \right]. 
\label{eq:1}
\end{equation}

\subsection*{Models and results}

We will consider three different models for the spectral data of the 29-keV doublet. For different values of the isomer transition energy $E_{\rm is}$ and the out-of-band branching $b$ of the 29.19~keV level into the ground state, we perform the maximal likelihood estimation of all other parameters of the considered model. We then obtain the value
\begin{equation}
X(\theta)=\sum_{i=1}^N \frac{(n_i-\lambda_i(\theta))^2}{\lambda_i}.
\label{eq:3}
\end{equation}
This value should be a $\chi^2_{N-p}$ random value, where $N$ is a number of points, and $p$ is the number of free parameters of the fit. To estimate the goodness of our fit, we calculate the {\em confidence level} $c(\theta)$, defined as the probability that a $\chi^2_{N-p}$ random value is larger than $X(\theta)$:
\begin{equation}
c(\theta)=P(\chi^2_{N-p}>X(\theta)).
\label{eq:4}
\end{equation}
The goodness of the model can be characterized by the goodness of the best fit, attained at optimal values of $E_{\rm is}$ and $b$. Now let us consider three different models and the corresponding results.

\subsection*{Model 1. Simple Gaussian peaks and linear background with a free slope}

\begin{figure}
\includegraphics[width=\columnwidth]{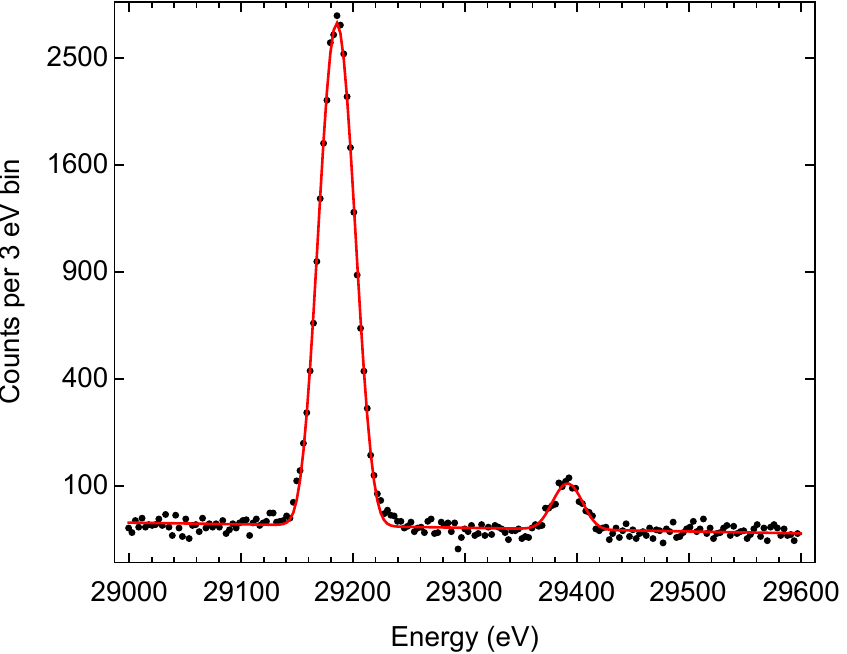}
\caption{Experimental data of Ref.~\cite{Beck2007eso} (black points) and the best fit of the free parameters corresponding to the model (\ref{eq:5}) (red curve).}
\label{fig:f1}
\end{figure}

Here we suppose that the expectations $\lambda_i$ are
\begin{equation}
\begin{split}
\lambda_i=\,& R_1 (1-b) e^{-\frac{(\EE-E_{\rm is}-E_i)^2}{2 \sigma^2}} + 
R_1 b e^{-\frac{(\EE-E_i)^2}{2 \sigma^2}}\\
&+R_2 e^{-\frac{(\EE-E_i+\Delta E_{42})^2}{2 \sigma^2}}+r_{\rm bg}-E_i \, s_{\rm bg},
\label{eq:5}
\end{split}
\end{equation}
where $\Delta E_{42}$ is a function of $E_{\rm is}$ defined in (\ref{eq:2}). This model contains the two parameters of interest $\{ E_{\rm is},b \}$, and six free parameters $\{\EE, \sigma, R_1, R_2, r_{\rm bg}, s_{\rm bg} \}$. The confidence level $c$ of this model nowhere exceeds 0.001, and we conclude that such a simple model is not valid. Moreover, the single-Gaussian model can be rejected at the 97.85\% confidence level ($c_{\rm max} = 0.0215$) even if $\Delta E_{42}$ is treated as a fit parameter. This means that the flaw of the model is not connected to any systematic errors of the energy calibration for the 29-keV and 42-keV regions of the spectrum, but points to a non-Gaussian shape of the instrumental response. The best fit corresponding to this model is shown in Fig.~\ref{fig:f1}.

\begin{figure*}
\includegraphics[width=\textwidth]{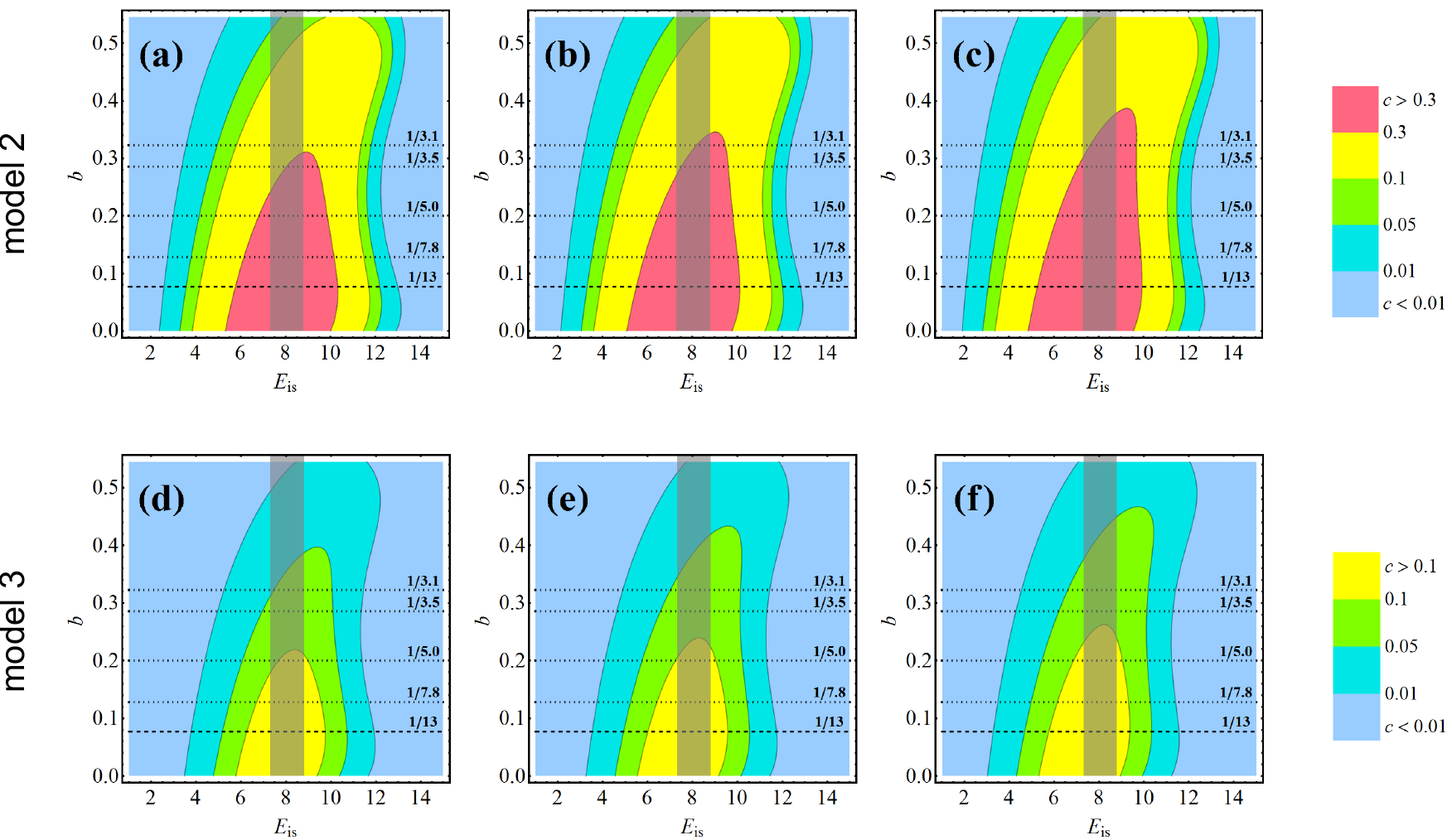}
\caption{Regions of various values of the confidence parameter $c$ (given in Eq.~\ref{eq:4}) (see the color scale on the right side for reference) corresponding to the maximal likelihood estimation of the free parameters within model 2 (given in Eq.~\ref{eq:6}, shown in the top row) and model 3 (Eq.~\ref{eq:6}, bottom row) on the $\{E_{\rm is},b\}$-plane for (a, d) $\Delta E_{\rm 42, meas}=198.22$~eV; (b, e) $\Delta E_{\rm 42, meas}=198.44$~eV, and (c, f) $\Delta E_{\rm 42, meas}=198.66$~eV. Here, the dashed horizontal lines corresponds to the value of the branching ratio $b=1/13$ \cite{Beck2007eso}, and dotted lines correspond to various values of $b$ meantioned in Refs.~\cite{Sakharov2010ote,Tkalya2015rla} and references therein. The shaded region corresponds to the range of $E_{\rm is}$ between 7.3 and 8.8~eV that was investigated in Ref.~\cite{Jeet2015roa}.}
\label{fig:contour}
\end{figure*}

\subsection*{Model 2. Double-Gaussian structure of peaks and linear background with free slope}

In Fig.~\ref{fig:f1}, one notes that 29.185-keV peak is a bit broader near its base than the best Gaussian fit. We speculate that during the operation time of the experiment, there were instances when the signal-to-energy calibration was insufficient to deliver the nominal resolution of about 26\,eV. The data obtained during these intervals has a larger spread, showing up as a broader Gaussian distribution. This hypothesis is introduced into our fit model by using a ``double Gaussian'' shape of the response function. Every monoenergetic line will be modelled as a superposition of two Gaussian functions with the same center position, but with different heights and widths. 

We introduce two new parameters into our model:~the relative heigh $B$ and the standard deviation $\Sigma$ of the broad component:
\begin{equation}
\begin{split}
\lambda_i=\,& R_1 \left[ (1-b) \left( e^{-\frac{(\EE-E_{\rm is}-E_i)^2}{2 \sigma^2}} + B e^{-\frac{(\EE-E_{\rm is}-E_i)^2}{2 \Sigma^2}} \right)\right. \\
& + \left. b \left(e^{-\frac{(\EE-E_i)^2}{2 \sigma^2}} + B e^{-\frac{(\EE-E_i)^2}{2 \Sigma^2}} \right) \right]\\
+&R_2 \left[ e^{-\frac{(\EE-E_i+\Delta E_{42})^2}{2 \sigma^2}}+B e^{-\frac{(\EE-E_i+\Delta E_{42})^2}{2 \Sigma^2}}\right]\\
+&r_{\rm bg}-E_i \, s_{\rm bg}.
\label{eq:6}
\end{split}
\end{equation}
Here, as before, we treat $\{ E_{\rm is},b \}$ as the parameters of interest, whereas we have $p=8$ free parameters: $\{\EE, \sigma, R_1, R_2, r_{\rm bg}, s_{\rm bg}, \Sigma, B \}$.  Model~(\ref{eq:6}) allows to perform a reasonably good fit in for a broad range of parameters; see Fig.~\ref{fig:contour}. Based on this model, a hypothesis about the absence of the isomer ($b=1$) can be rejected at a level of 99.985\% within the $\pm 3 \sigma$ range of $\Delta E_{42}$.

\subsection*{Model 3. Double-Gaussian structure of peaks and linear background with fixed slope}

Generally speaking, the background counts may be produced by various processess. However, we can expect that $\gamma$-particles changed their energy in Compton processess within the source and constructions surrounding the detector give the main yield into the background. Let us suppose, for the sake of simplicity, that the spectrum of these gammas is energy-independent between 29.0 and 29.6\,keV. Then the slope of the background count rate appears due to the variation of the stopping power of the detector material. If this variation is smooth enough to be approximated by the linear function, the coefficient $s_{\rm bg}$ in (\ref{eq:6}) is not a free parameter, but takes a form $s_{\rm bg}=r_{\rm bg} \times d_{\rm bg}$, where $d_{\rm bg}$ is determined by the properties of the absorber. This results in a model
\begin{equation}
\begin{split}
\lambda_i=& R_1 \left[ (1-b) \left( e^{-\frac{(\EE-E_{\rm is}-E_i)^2}{2 \sigma^2}} + B e^{-\frac{(\EE-E_{\rm is}-E_i)^2}{2 \Sigma^2}} \right)\right. \\
& + \left. b \left(e^{-\frac{(\EE-E_i)^2}{2 \sigma^2}} + B e^{-\frac{(\EE-E_i)^2}{2 \Sigma^2}} \right) \right]\\
+&R_2 \left[ e^{-\frac{(\EE-E_i+\Delta E_{42})^2}{2 \sigma^2}}+B e^{-\frac{(\EE-E_i+\Delta E_{42})^2}{2 \Sigma^2}}\right]\\
+&r_{\rm bg}(1-d_{\rm bg}\, E_i),
\label{eq:7}
\end{split}
\end{equation}
where the coefficient $d_{\rm bg}$ is a property of the detector material.

Beck \textit{et al}.~have used a NASA X-ray spectrometer \cite{Porter2004tae} whose absorber is made of ${\rm 8~\mu m}$ thick HgTe alloy. Taking the data on the transmission of such an absorber from the database \cite{Xray}, and assuming an energy-independent spectrum of scattered $\gamma$-particles, we find that the background count rate is almost linear between $29.0$ and $29.6$~keV; see also Fig.~\ref{fig:fig2} for a broader range. The slope can be characterized by a coefficient $d_{\rm bg}=2.44312 \times 10^{-5}\,{\rm eV^{-1}}$.
 
Fixing the $\{E_{\rm is},b\}$-pair, we have 7 free parameters of the model (\ref{eq:7}): $\{\EE, \sigma, R_1, R_2, r_{\rm bg}, \Sigma, B \}$). The fit occurs to be not as good as for the previous model, which is not surprising: as can be inferred from Fig.~\ref{fig:fig2}, the background does not drop off exactly linearly with energy. Still, the boundaries of the confidence regions resemble those of the prevous model (\ref{eq:6}); see Fig.~\ref{fig:contour}. The reduction of the confidence level may result from the incorrectness of our hypothesis about an energy-independent spectrum of the background $\gamma$-particles, and/or from some different processes contributing to the background. 

\section{Confidence region based on the lineshape of the 29.185-keV line}

The confidence regions constructed in the previous section rest on two assumptions, namely (1) that the energy calibration of the detector is correct and does not introduce a significant systematic error, and (2) that the determination of $\Delta E_{42}$ was correct.

(1) Concerning the energy calibration, we estimate that an error of a few eV in the position of the lines used for calibration does not change the value of $E_{\rm is}$ by more than a few 0.1\,eV. We do note, hoewever, that some of the calibration lines are spaced by 10\,keV and do not capture detector non-linearities on smaller energy scales.

\begin{figure}
\includegraphics[width=\columnwidth]{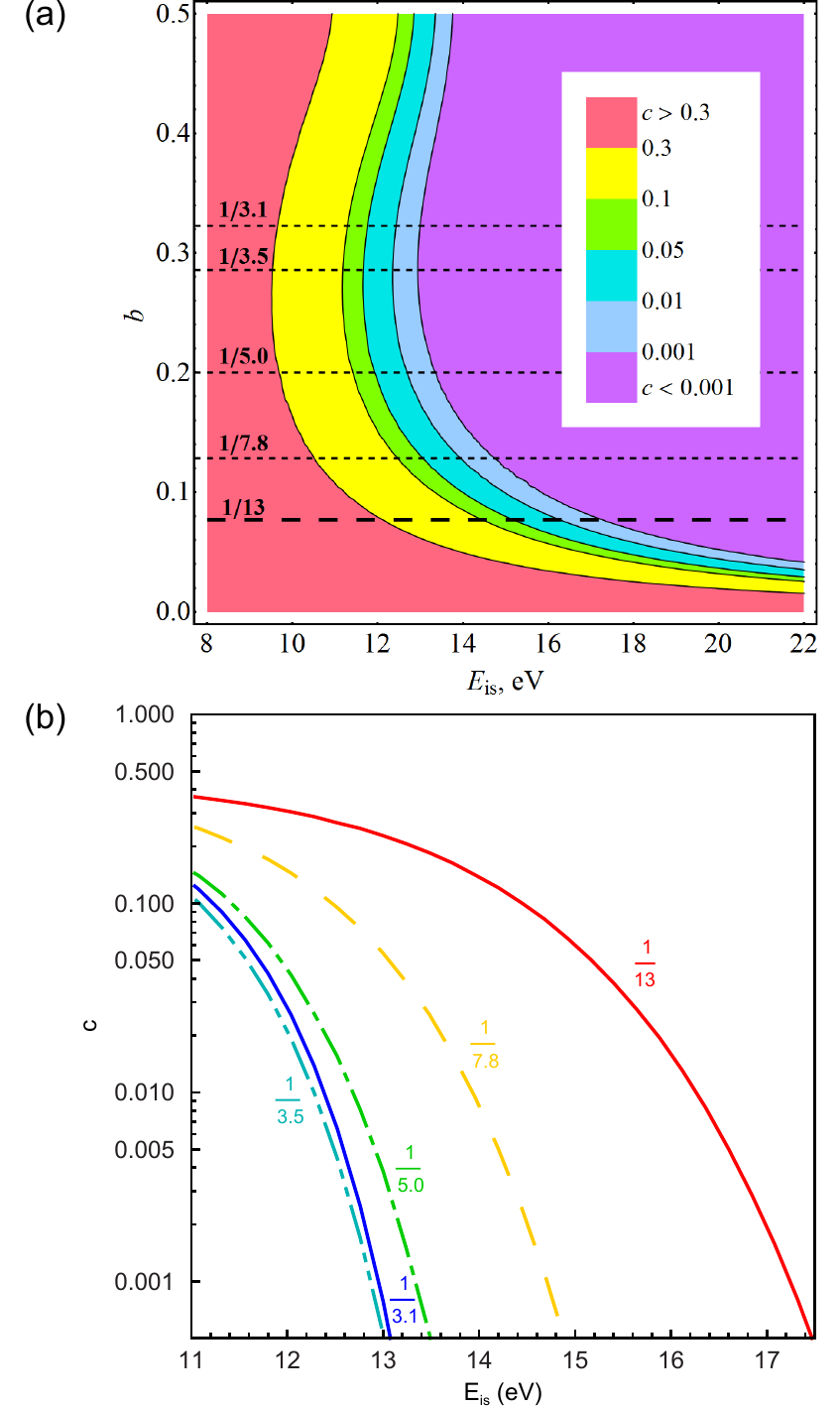}
\caption{Confidence region of $\{E_{\rm is},b\}$ constructed solely from the 29.185\,keV peak. (a) Contour plot showing $c$, and (b) cut through the plane at various assumed values  of $b$, corresponding to the horizontal lines in (a).}
\label{fig:lineshape_only}
\end{figure}

(2) The original figure in Ref.~\cite{Beck2007eso} shows the data of only one out of 25 pixels in the 42-keV region, thus does not provide the full data set required to perform a full re-analysis. As in Ref.~\cite{Beck2009ivf}, we assumed the out-of-band branching ratio to be $b_{42}=1/50$ throughout the present work. Uncertainties concerning this value have been addressed in Refs.~\cite{Sakharov2010ote,Tkalya2015rla}.

To bypass these two assumptions, we construct a confidence region based solely on the lineshape of the 29.185\,keV feature; see Ref.~\cite{Kazakov2014pfm} for a similar treatment. Very similar to model 2 (\ref{eq:6}), we fitted all three peaks in the 29-keV spectrum, but left $\Delta E_{42}$ as a fit parameter. A contour plot is shown in Fig.~\ref{fig:lineshape_only}(a), and a graph assuming various values of $b$ is shown in (b). To give two examples, for $b=1/13$, $E_{\rm is} > 15\,$eV can be excluded at the 95\% confidence level, and for $b=1/5.0$, $E_{\rm is} > 12\,$eV can be excluded at the 95\% level as well.

This analysis might be valuable for experiments with limited tolerance towards larger-than-expected deviations of $E_{\rm is}$ from the currently accepted value of 7.8(5)\,eV, \textit{e.g.}~Th$^+$ ion traps (second ionization energy at 11.9\,eV) and doped crystals (VUV transmission cut-off around 10\,eV).


\bibliographystyle{apsrev}
\bibliography{bib}

\end{document}